# Falsification or Confirmation: From Logic to Psychology

**Roman Lukyanenko**
Information Systems Department
Florida international University
rlukyane@fiu.edu

**Abstract**

Corroboration or Confirmation is a prominent philosophical debate of the 20th century. Many philosophers have been involved in this debate most notably the proponents of confirmation led by Hempel and its most powerful criticism – the falsification thesis of Popper. In both cases however the debates were primarily based on the arguments from logic. In this paper we review these debates and suggest that a different perspective on falsification versus confirmation can be taken by grounding arguments in cognitive psychology.

**Introduction**

The idea of scientific justification, a key concept of the philosophy of science, has been a common ground of intellectual battles of the 20th century. Among many others, corroboration advocated by Karl Popper and verification presented by Carl Hempel attempted to provide contrasting accounts of scientific progress. Using formal logic as a theoretical foundation, Hempel and Popper were concerned with both rationalizing as well as developing an optimum method of scientific discoveries. While both falsification and verification provide account of scientific reason, they contain a number of contradictions and ultimately fail to capture the full complexity of a scientific process.

Both verification and falsification contain powerful arguments that support their models. Verification stems from human's desire to quantify a probability of a similar event occurring again in the same way. Ensuring this probability is a theory that is derived from a number of observations. Major proponent of verification, Hempel argued that each new confirmation strengthens the theory, and diversity of experimentation is important. Some theories cannot be tested directly, and bridge concepts become vital as they extend theoretical articulations into observable realm. Simplicity is what makes, according to Hempel, one theory better than another.

From the logical point of view (as opposed to cognitive that is discussed later), verification fails due to a problem of induction that has been identified by Hume and later adopted by Popper. An induction moves from a statement of a narrow scope to a statement of a broader scope. In an inductive argument, conclusion always goes beyond the original premises:

> H1. I observed a thousand crows, and they were black.
> H2. Therefore, all crows are black.

Pondering on the mechanisms of induction, Hume reasoned, inductive statements are impossible to justify. One can not apply deductive logic to it, and similarly, cannot use induction itself to prove it without inevitable circularity. As basic form of reasoning, Hume concludes, inductive statements cannot be justified (Harris 1997, p. 52). Adopting Hume's conclusions, Popper, argued further that experiments can only produce a one to one statement, and cannot generalize – a process necessary for scientific advancement.

Responding to verificationists that focused on experimentation, Karl Popper argued that fundamentally, science begins with problems and questions, and not with experiments (Popper 2003, p. 222). Experimentations, he reasoned may occasionally give rise to theories, but only if they contradict, or "clash" with existing theoretical beliefs. Ultimately, he continued, science seeks solutions and understanding, and both can only be found in theories, not experiments. This reasoning suggested a deductive account of science.

In the domain of logical reasoning, verification contains an internal contradiction, which was exploited by Popper as one of the main justification behind an alternative deductive method. Popper noticed that from a logical perspective, verification is regressive. Hempel's primary measurement of theory's acceptability is the principle of simplicity. Assuming that "the basic laws of nature are simple" (Hempel 1966, p. 42), Hempel regarded simple theories as more probable. Popper, employing a basic premise of probability: probability of one is always more than probability of sums, noticed that on the basis of logic, one verification is probabilistically stronger than several. Thus, the core of verificationism that Hempel developed into concepts of "diversity of evidence" (p. 35) and confirmation by new evidence (p. 37) have been undermined.

Simplicity of the verification theory also fails because it necessitates a need for simplicity measurement. As Hempel (1966) remarked "any criteria of simplicity would have to be objective," but he then admitted that "it is not easy to state clear criteria of simplicity" (p. 41). Yet, such even if it achieves a maximum objectivity, itself will have to be a theory. Therefore, the principle of simplicity inevitably leads to a circular argument: one theory will be used to verify / measure another, ultimately producing an infinite logical loop.

An alternative to verification according to Popper was falsification or corroboration of theories. The logic of scientific justification for Popper is a fundamentally deductive one. Theories are born as guesses and the role of science is to test them in order to reveal their falsities. The theories that withstand the tests become corroborated, while the ones that fail are called falsified. Failures, however, are important stepping stones of progress, as people learn from their own mistakes.

Unlike induction that logically cannot offer truth, falsification is truth apt due to its deductive foundation. Yet, Popper realizes that universal truth is unattainable. To overcome the inherit limitation of verification, Popper offered a concept of *verisimilitude,* "approximation of truth" (Popper 2003, p 234). Verisimilitude suggest that even the most brilliant theories are "at best approximations" (Popper 2003, p 234).

Popper's falsification account too has internal, logical contradictions. While deductive logic is knows to be truth-apt due to its form (Hacking 1983), it may contain generalizations about reality that cannot possibly be verified. It is logical, as Popper suggests deriving a statement of a narrower scope, yet, the original statement remains forever unproven. This logic leads to an inevitable conclusion, that in a strict account of deductive logic, no scientific knowledge can ever be verified.

A commonly accepted weak point of falsification is the auxiliary theory argument. Given imprecise measurements, techniques, changing subjects etc, it is possible to defend a "falsified" theory by either adjusting it or adding a new condition that would void the falsification. This, in turn, would cause a theory to grow and become unnecessarily complex.

Falsification if applied staunchly can cause many progressive theories to be rejected before they have an opportunity to flourish. Copernucus model of the solar system contradicted the "observations" of his time. Einsten's theory of general relativity contradicted well established laws of physics and based on them common practice observations. By the logic of strict falsification, Einstein's theory should have been rejected. Interestingly, Popper himself was known to be Einstein's intellectual admirer.

One of the limitations of both Hempel's verifications and Popper's falsification theory is their over reliance on logic as a fundamental form of reasoning. Work in cognitive psychology shows that many processes outside of the logical domain account for much of our thinking, and therefore of the endeavor to extend our understanding of reality. A concept of similarity, for example, as one of the basic cognitive processes (Rumelhart and Abrahamson 1973; Imai 1977), has seen by cognitive scientists as one of the driving forces behind scientific discovery. Similarity judgment is ubiquitous and is closely linked with such cognitive processes as classification, concept formation, and generalization (Rumelhart and Abrahamson 1973; Imai 1977, Tversky 1977). According to cognitive sciences, powerful and sometimes subtle processes of analogy, generalization, specialization play significant role in scientific *justification* (Induction 1986, p. 288, p. 326), the concept rejected by logical empiricists. Responding to Reichenbach, and indirectly to Popper, psychology argues that the distinction between "context of discovery" and "context of justification" is a superficial one, as there is an "intimate connection between…*how* science is done…and how *ought* to be done" (Induction…1986, p. 320). Extending the cognitive argument, we see that logic is one of the many tools of scientific justification, and therefore, logical contradiction of, for example problem of induction, while valid within own domain, cannot account for a full complexity of human's quest to understand reality expressed through science. While controversial on its own grounds, cognitive psychology went as far as claiming that "analogy is the primary means of theory construction" (Induction…1986, p. 326). While this approach itself is limiting, it does point to the integrated view of the context of discovery" and "context of justification," which allows us to see that both formal deduction of Popper and induction of Hempel work well only if phenomena can be easily observed. Yet, many problems of science deal around abstract, non-observable issues. In situations like that, justification often has to come from the context of discovery and use analogies, conceptual combinations, and mixture of inductive and deductive reasoning.

    The reliance on logic and probability has been questioned by a number of experiments that reveals a disconnect between human actions and logical and probabilistic norms.

Representativeness, common group of "systematic errors" in human cognition (Tversky and Kahneman 1974), reflects the probably of one concept belonging to another, broader concept. Representativeness has a number of judgment biases that pose a challenge to the logical account of scientific justification. "Insensitivity to prior probabilities of outcome" describes an error when prior given or known probabilities are ignored. A description that is stereotypical of a lawyer will most likely be labeled as the description of a lawyer, even when respondents are aware that the number of lawyers is very small in the given sample (Tversky and Kahneman 1974, p. 1125). "Insensitivity to sample size" reveals the failure to consider sample size in predicting outcomes. It is based on the mistaken belief that "chance is … a self-correcting process in which a deviation in one direction induces a deviation in the opposite direction to restore equilibrium" (Tversky and Kahneman 1974, p. 1125). In reality deviations are diluted through large sample size. "Insensitivity to predictability" demonstrates tendency to see future similar to the present situation, even though the later may present few clues to the future events (Tversky and Kahneman 1974, p 1126). People select outcomes that are more similar to the inputs. The more consistent the inputs, the more confident people are of the outcomes. Yet, statistically, redundancy among inputs decreases accuracy. These inherit biases proliferate into our regular and scientific thinking, and often compete with inductive and deductive forms of logical reasoning.

      Scientific justification and important concept for understanding scientific method and creating prescription for scientific progress has been approached by both traditional verification and falsification accounts as a logical program. Within the logical framework, Popper and Hempel provided contrasting account of scientific reason. While each found sound

contradictions of opposite theories, they failed to capture the full complexity of a scientific process which comes from broadening the perspective on the concept of scientific justification.